# Flight and ground demonstration of reproducibility and stability of photoelectric properties for passive charge management using LEDs

S. Buchman[1], T.S.M. Al Saud[2], A. Alfauwaz[3], R.L. Byer[1], P. Klupar[4], J. Lipa[1], C.Y. Lui[5], S. Saraf[5], S. Wang[5,6], P. Worden[4]

[1]Stanford University, W.W. Hansen Laboratory, Stanford, US
[2]Saudi Space Commission, Riyadh, Saudi Arabia
[3]King Abdulaziz City for Science and Technology, Riyadh, Saudi Arabia
[4]Breakthrough Initiatives, Moffett Field, US
[5]SN&N Electronics Inc., San Jose, US
[6]Hainan Tropical Ocean University, Sanya, China

Charges as small as 1 pC degrade the performance of high precision inertial reference instruments when accumulated on their test masses (TMs). Non-contact charge management systems are required for the most sensitive of these instruments, with the TMs free-floating, and their charges compensated by photoelectrons in a feedback loop with a TM charge measurement. Three space missions have successfully demonstrated this technique: the Relativity Mission, Gravity Probe-B (GP_B) launched in 2004, the LISA Pathfinder (LPF) launched in 2015, and the UV-LED mission launched on SaudiSat 4 in 2014; with the first two using the 254 nm Hg discharge line and the last one a set of 255 nm UV-LEDs. UV-LEDs represent a significant improvement over the discharge sources, in terms of reliability, lifetime, switching speeds, power consumption, weight, and volume. Charge management techniques that eliminate the charge measurement and feedback systems, referred henceforth as passive, reduce the complexities and disturbance effects introduced by these systems, and are thus the subject of active research and development work. Passive charge management depends critically on the stability and reproducibility of the photoemission properties of a given system. In support of this work, we present comprehensive flight characterization data for a suite of 16 UV-LEDs in various configurations and $255 \pm 1$ nm center wavelength. Flight data was acquired between December 2014 and December 2015 with the UV-LED instrument flown on SaudiSat 4. We back up our results with ground-based measurements performed in configurations comparable to the flight one between September 4, 2020, and October 8, 2020. All results confirm the excellent reliability of the UV-LEDs in the space environment, are fully consistent with the findings of ground studies, and support the approach of using LEDs for passive charge management. We find that the equilibrium potential of the TM, under illumination by the 255 nm LEDs, is independent of the UV intensity and reproduceable to about $\cong \pm 6$ mV, or $\pm 6$ fC/pF, over periods of up to six months. The value of the equilibrium potential is dependent on the geometry of the electric field between the TM and its enclosure, and thus on the exact configuration of the instrument.

## I. Introduction

To date, the best performance for inertial sensors in the space environment has been achieved using bulk test masses (TMs) for missions like DISCOS[1], GRACE[2], GOCE[3], GP-B[4], the pathfinder LPF[5] for the LISA[6] array, and MICROSCOPE[7]. All these instruments are designed to minimize both external and internal disturbances using a dedicated TM, a housing for the TM, EMI shielding, a system for nulling of the local gravitational field gradient, and various auxiliary systems. Main internal disturbances are caused by gravity gradients from the instrument and instrument platform, residual gas in the housing, temperature variations and gradients, and electromagnetic interactions between TM and its environment[8,9]. External disturbances are due to radiation pressure, electromagnetic fields, platform propulsion, and residual environmental gas. Electric charges on the TM interact with internal and external electromagnetic fields, degrading the performance of the instrument, and are therefore limited by performance requirements. Examples of TM charge limit requirements are $\leq 15$ pC for GP-B[10,11] ($C_{GPB} \cong 1$ nF) and $\leq 3$ pC for LPF and LISA[12,13,14].($C_{LPF/LISA} \cong 34$ pF).

In this paper we present satellite flight data, backed up by ground experiments, that validate the use of UV-LEDs for the technique of passive charge management (PCM)[15], (known also as self-adaptive[16]), a significantly simplified approach, that will improve instrument performance while significantly reducing mass and complexity. The PCM techniques has been developed specifically for instruments with spherical TMs, TM to housing gaps larger than 1 cm, and no electrostatic sensing or activation systems. Systems with capacitive sensing and electrostatic activation that have variable and large internal electric fields, LISA for example, will require special adaptations of any PCM method.

Two approaches have been proposed for the non-contact charge management of TMs: UV generated photoelectrons[10] and electrons generated by field emission cathodes[17], with the former adopted as baseline for space applications. GP-B[18] and LPF[19] have successfully demonstrated charge management for space-based inertial sensors using photoelectrons generated by the 254 nm UV line of Hg discharge lamps combined with the force modulation method of TM charge measurement[10]. A third method using ionized gas[20], has been implemented for the suspended mirrors of the ground-based Laser Interferometer Gravitational Observatory (LIGO)[21].

The development of LEDs in the range of wavelength at and below 255 nm[22], has prompted the replacement of Hg lamps with LEDs as UV sources for the charge management of space-based inertial sensors. LED sources are much lighter and have much lower power consumption than Hg lamps[23]. Furthermore, LEDs can be modulated at frequencies well above the bandwidth of the inertial sensors, thus reducing disturbances by being compatible with AC charge management[24]. A collaboration of Stanford University, NASA and KACST[25] has successfully flown a technology demonstration of charge management with UV-LEDs[26], the UV-LED Mission, on the SaudiSat 4 small satellite[27] launched in June 2014. Flight data presented in this work are acquired from this mission.

For PCM to work, the reflectivity and the photoemission efficiency of the system are required to be stable for periods long compared to the days to weeks typical time for the TM to reach its maximum allowed charge from cosmic radiation, and repeatable to levels corresponding to less than 10% of this charge i.e., $\leq 0.3$ pC ($\cong 9$ mV for $C_{LPF/LISA}$). The UV-LED Mission flight data presented validates these requirements, and thus the use of UV-LEDs for space applications in general and for the PCM method in particular. The TM and its housing have similar gold coatings and the gap between them is flooded with between 0.13 mW and 1.0 mW of UV light. Ground testing in systems with similar configurations to the flight ones has validated the flight results[15].

The paper organization is as follows. Section II contains a description of the flight instrument. In sections III and IV we describe respectively the flight and ground experiments and their results that demonstrate the reproducibility to $\cong \pm 6$ mV ($\cong \pm 0.2$ pC for 28 pF) of the TM charge under 255 nm UV illumination, for our system configuration, and 0 V TM enclosure bias. These results validate with space data two proposed PCM approaches: a) single source low energy photoelectrons[15] and b) balanced dual source high energy photoelectrons[15,16]. Discussion of results and conclusions are given in sections V and VI.

## II. Flight Instrument

Details of the flight mission are given in reference 26. Here we provide a detailed analysis of the flight results focused on their applicability to the PCM method.

Figure 1 shows a photograph of the science instrument for the mission with axis notations. The instrument consists of an 8.9-cm diameter TM gold coated hollow aluminum sphere, a parallelepipedal housing with four bias plates, two experiments containing eight UV-LEDs each, and the control and measurement electronics. In the flight instrument the vertical *x* and *y* inner surfaces are connected and serve as bias for the charge management, while the horizontal *z* surfaces are grounded[26]. The digitization bin, $\delta V_{AD}$, of the readout of the potential is $\delta V_{AD} = 2.4$ mV.

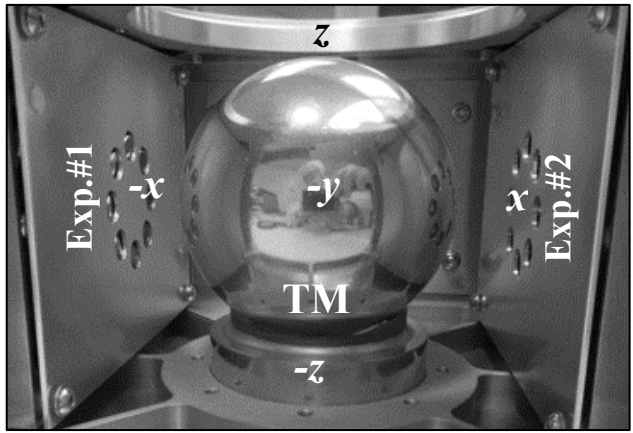

**Figure 1. Open view photograph of instrument**

Total capacitance of TM to its housing is $C = 28$ pF, corresponding to a TM potential of 36 mV for a TM charge of 1 pC. Note that LPF and LISA have a similar capacitance of 34.2 pF[19]. The instrument has been optimized for PCM by coating both TM and its housing with the same material – in this case gold. We find that the work function of the deposited Au film is centered at about 4.6 eV[15] and spans from 4.3 eV to 4.9 eV[28]. Note that the values for the Au work function differ significantly by author: 4.3-4.9 eV[29], 4.86 eV[30], 5.38 eV[31], 5.31-5.37-5.47 eV (for the 111, 110, and 100 crystal orientations respectively)[32], 5.1 eV[33], 4.83±0.02 eV[34,35]. We are observing the decrease to a minimum of about 4.3 eV in the work function, as obtained by some authors after contamination by exposure to air[36].

Figure 2 left is the CAD model of the electronic boxes, the TM, and the insulation tubes holding the charge amplifiers and Figure 2 right is a CAD of the electronic boards of the experiment showing: TM, wire routing, and holding tubes, (with payload structure removed). Most disturbances of the TM by its housing scale with the inverse second or third power of the gap size between TM and housing[8], making larger gaps - of the order of the TM radius - appropriate for best performance. In this system 'nominal/average' distance of TM to housing gap is about 3 cm. However, charge measurement becomes more difficult as the gap is increased to the size of the TM, 1 cm and above, as its sensitivity also scales as the inverse third power of the gap[10]. All inertial sensors, and in particular those not requiring electrostatic actuation (spherical TM for example) would greatly profit from a charge management system that does not require charge measurement and can be activated over short periods at a frequency well above the operational band of the inertial sensors.

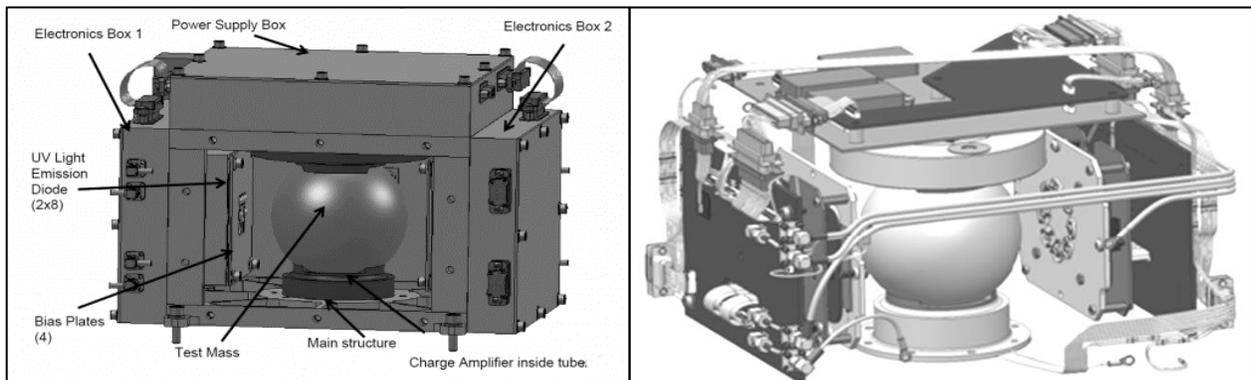

**Figure 2. CAD models: (left) electronic boxes, TM, and insulation tubes holding charge amplifiers (right) experiment boards, TM, wire routing, and holding tubes with payload structure removed.**

Figure 3 is a photograph showing the LEDs of the UV-LED mission for Experiment #1 (left) and Experiment #2 (right). Experiment #1 consists of 8 LEDs mounted in TFW-TO39 canisters (4 with hemispherical-lens windows and 4 with tall windows), while Experiment #2 has 8 surface-mounted LEDs (4 with flat windows and 4 with ball-lens windows). All LEDs are aligned so as to point to the center of the TM. For further technical details see references 28 and 42. Two amplifiers, connected directly by push pins to the top and bottom of the TM ($\pm z$ axis), measure redundantly its potential.

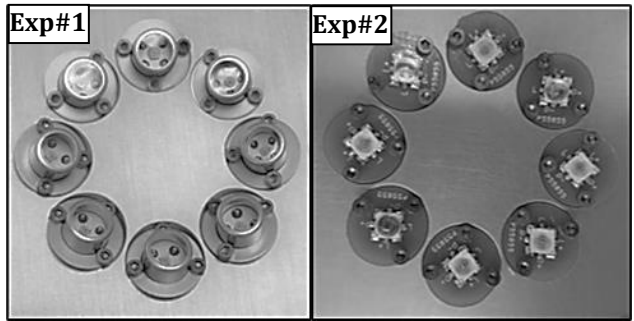

Figure 3. The LEDs on Experiment #1 (left) and Experiment #2 (right)

Film samples, similar to the thin films on the TM and the bias plates, were deposited from a 99.9% gold sputtering target. A gold film thickness of 1000 Å was chosen by requiring that the ratio of photoelectrons generated at the interface between the gold coating and its substrate to the photoelectrons generated in the gold film be small[28]. Given the 70 Å electron mean free path in gold[29], this ratio is $< 10^{-6}$. We define the photoemission (or quantum) efficiency of the coatings, $\Psi$, as the ratio of the emitted photoelectron flux, $\Phi_{emitted}^{electron}$, to the total incident photon flux, $\Phi_{incident}^{photon}$:

$$\Psi = \Phi_{emitted}^{electron}/\Phi_{incident}^{photon}; \quad \Psi = (hcI_{photoelectric})/(e\lambda P_{LED}) \tag{1}$$

where $I_{photoelectric}$, $P_{LED}$, $h, c, e,$ and $\lambda$ are the photoelectric current, the UV-LED power, the Planck constant, the velocity of light, the electron charge, and LED wavelength.

For the 255 nm LEDs, an incident beam of power $P_{photon} = 50\mu W$ generated a current of $I_{electron} = 3.49$ pA for our film samples[28], resulting in a value of $\Psi = (3.40 \pm 0.35) \times 10^{-7}$, somewhat low compared with other work[29-35]. Tests with similar gold coatings performed in a system with better vacuum and more comparable geometry gave[15]: $\Psi = (5.9 \pm 0.2) \times 10^{-7}$, the value we will use henceforth. The reflectivity coefficient is the ratio of reflected photons to incident ones: $R \equiv P_{reflected}/P_{incident}$. For the film samples[28] $R = 0.17$ at 255 nm - somewhat low with respect to other measured values: $R = 0.28 \pm 0.04$[37] [38] and $R = 0.33$[39] [40]. This low value of R was however validated for our gold films in tests by S. Wang et. al.[15] (in the same test series as for the above $\Psi$ measurement) as being: $R = 0.18 \pm 0.01$, the value used hereafter.

The LEDs are AlGaN devices of different configurations, specially designed with integrated photodetectors (PDs) by Sensor Electronic Technology[41], with center frequency $\lambda = 255 \pm 1$nm and full width at half maximum of $FWHM = 11 \pm 1$nm. In energy units: $E_v = 4.86 \pm 0.02$eV, $E_{FWHM} = 0.21 \pm 0.02\,\text{eV}$. The LED efficiency, $E_{LED}$, is defined as $E_{LED} \equiv P_{LED}/I_{LED}$, where $I_{LED}$ is the LED excitation current and the manufacturer specification is $E_{LED} = 15\,\mu\text{W/mA}$. Optical power, measured at our conservative space-operations current of 10 mA, was $160 \pm 10\,\mu\text{W}$ for the typical 255 nm LEDs[42], in agreement with specifications. Figure 4 shows, from top to bottom, three of the four LED configurations used: TO39 header mounts with hemispherical lens windows, TO39 with flat windows, and a surface mount with flat window. The surface mount with ball-lens windows not shown. The calculated photocurrent from the TM, by direct illumination versus LED power, $(I_{TM}^D)_P^C$, or current, $(I_{TM}^D)_I^C$, for one LED at 100% duty cycle, are given by:

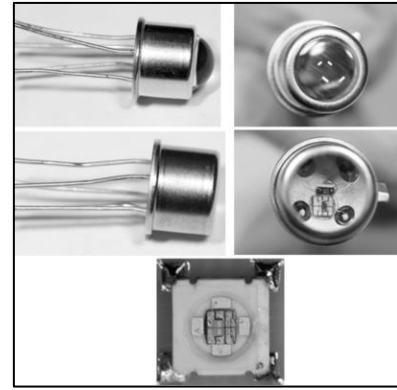

Figure 4. UV-LED packages. From top: TO39 hemispherical lens, TO39 flat window, surface mount with flat window

$$[(I_{TM}^D)_P^C]/P_{LED} = \frac{e\lambda}{hc}\Psi \cong (120 \pm 4)\,\text{fA}/\mu\text{W};\quad [(I_{TM}^D)_I^C]/I_{LED} = \frac{e\lambda}{hc}\Psi E_{LED} = (1.9 \pm 0.1)\,\text{pA/mA} \quad (2)$$

Henceforth we will calculate all $I_{TM}$ values as function of $P_{LED}$, that is given in fA/µW units.

Multiple reflections can be neglected as their contributions to $\Phi^n$, the sum of second and higher reflection number flux, as ratio of $\Phi^0$, the initial total flux to the TM or its housing, is given by:

$$\Phi^n/\Phi^0 < \sum_{n=1}^{\infty} R^{2n} = R^2/(1-R^2) \cong 0.03 \quad (3)$$

and therefore, below the experimental errors – with $n = 1,2,3\ldots$. Further reduction in the reflected flux is due to the geometry of the system. A fraction of the photons are not returning to the TM or its housing after two reflections causing the contribution to the photocurrent to be less than that given in equation 4.

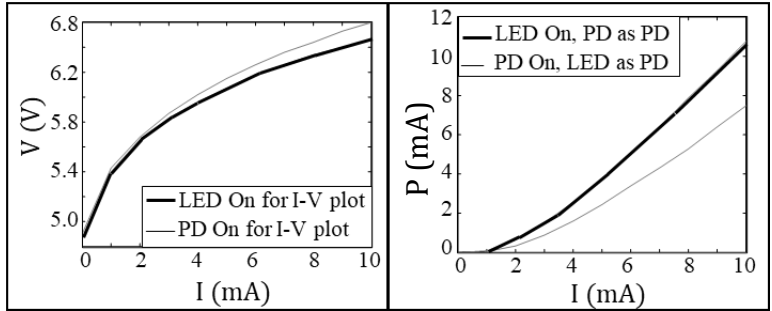

Figure 5 V vs I (left) and P vs I (right) curves showing interchangeability of nominal LED-UV and PD.

Each LED has its own PD used to monitor power output. In our devices, the PD is a second LED on the same die, and UV-LED and PD are interchangeable as performance. Figure 5 shows the close correspondence between the performances of the UV-LED and the PD, where the heavy traces correspond to the nominal UV-LED powered on and the PD used to read out the UV power and the thin lines correspond to the inverted usage; with voltage vs current on the left panel and power, in terms of photodetector current, vs current on right one. The nominal PD sensitivity is $I_{PD}/I_{LED} \approx 1\,\text{mA/mA} \Rightarrow P_{LED}/I_{PD} \approx 63\,\mu\text{W/mA}$ for LED currents $\geq 4$ mA at 100% duty cycle.

Environmental and radiation sensitivity testing was performed prior to flight[28, 43]. Each of the two experiments is run separately, while the four bias plates have a common bias voltage. A detailed description of the instrument and the experimental procedure is given in references 26 and 28. Figure 6 shows the typical performance of a flight UV-LED before and after environmental testing: thermal-vacuum testing and vibration-shake-drop testing, with LED current vs voltage (upper left), LED optical power vs voltage (upper right), PD current vs LED current (lower left)

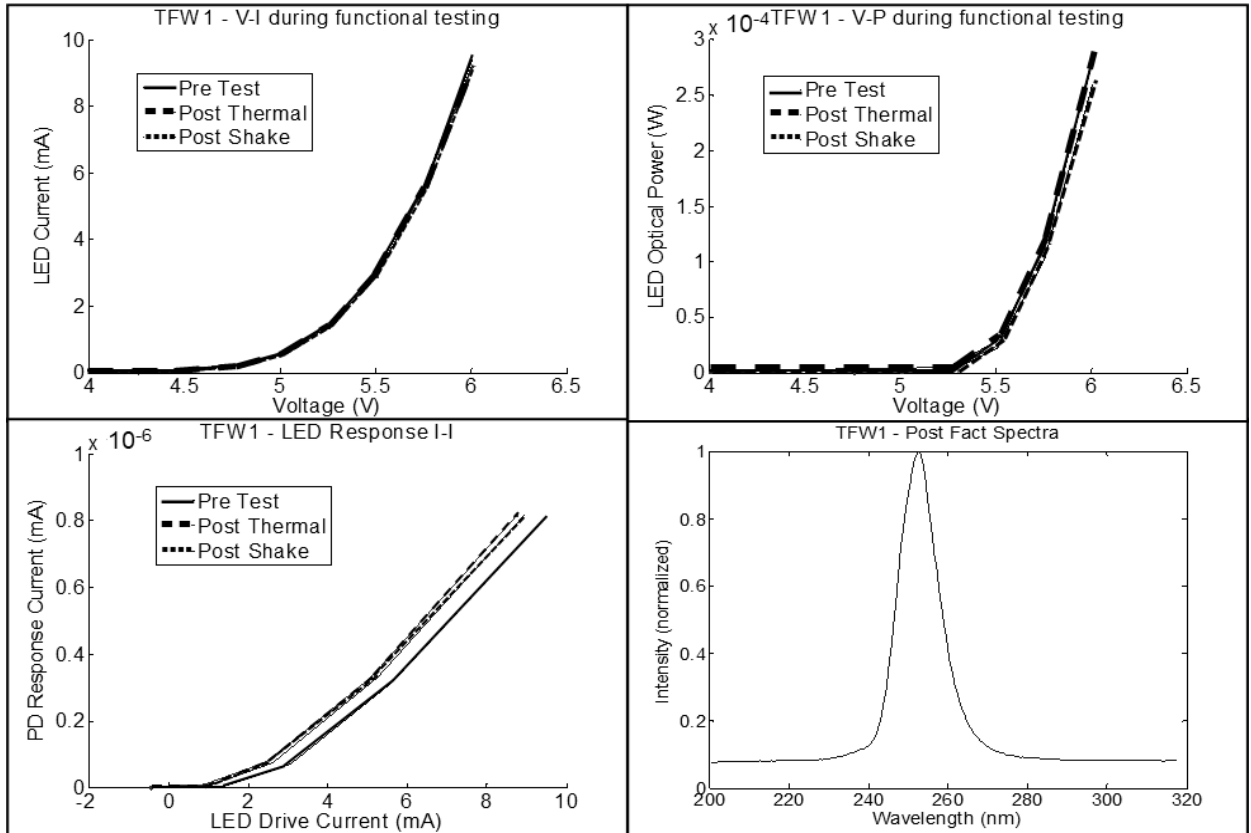

**Figure 6. UV-LED performance pre and post environmental testing: I vs. V (up left), P vs. V (up right), $I_{PD}$ vs. $I_{LED}$ (down left), power spectrum (down right)**

and LED spectrum as normalized power intensity vs wavelength (lower right). Reference 26 also gives detailed data for the performance of the UV-LEDs at the beginning of the mission and after 12 months on orbit. No significant change is observed for all key UV-LED performance parameters after one year in orbit. Figure 7 shows the optical power of the LEDs, measured in space by the PDs as function of their excitation current, with 20% duty cycle on 7/22/2015: Experiment #1 left panel and Experiment #2 right panel. The range of powers between LEDs varies by about ≈ 20% for Experiment #1 and ≈ 7% for Experiment #2 are:

$$\langle I_{PD}/I_{LED} \rangle^{Exp1} \approx (0.10 \pm 0.02) \text{ mA/mA}, \quad \langle I_{PD}/I_{LED} \rangle^{Exp2} \approx (0.23 \pm 0.02) \text{ mA/mA} \quad (4)$$

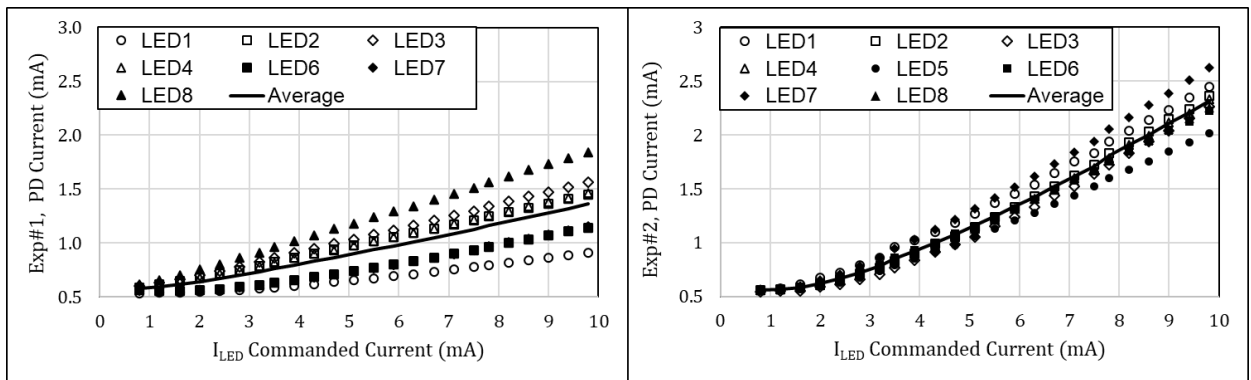

**Figure 7. Power versus excitation current 7/22/2015: Exp#1 left panel, Exp#2 right panel**

consistent for Experiment #2 with the ground measurements for a single LED shown on Figure 5 right and Figure 6 lower left, $I_{PD}/I_{LED} = (0.20 \pm 0.02)$mA/mA for 20% duty cycle. Experiment #1 produces about half the optical flux of Experiment #2. Note, that the ground data was measured with a calibrated power meter, while the calibration of the PDs is less accurate.

### III.a. Flight Experimental Results, General Considerations

LISA has the most stringent requirement for maximum TM charge allowed[12,13], $Q_{TM}^{max} \leq 3$pC, translating into $V_{TM}^{max} \leq 100$ mV for our UV-LED instrument. For LPF and LISA the maximum allowed charging results in a TM voltage of $\leq 80$ mV. Seen that cosmic radiation charging for LPF and LISA has been shown by calculation[11,44] and confirmed by flight data[13] to be positive, experiments have mainly focused on the discharge of the TM from positive potentials.

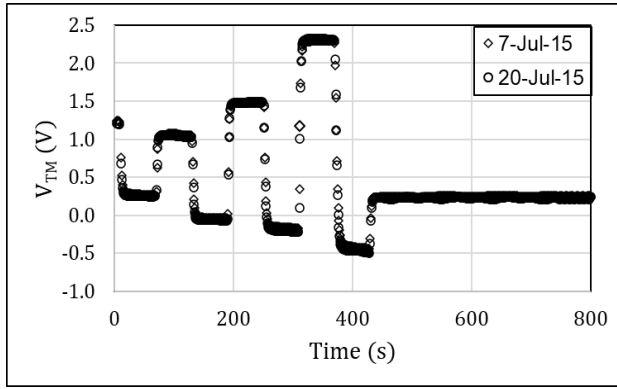

**Figure 8. Repeatability of charge control data**

To validate the use of UV-LEDs for PCM we demonstrate that the equilibrium TM potential is independent of the amplitude of the photoelectric current and consistent over time. Passively reducing the TM charge to $\leq 0.1 \times Q_{TM}^{max}$ will require either photoelectrons with energy only slightly above the work function of the coating material of the TM and its housing[15], or balancing the photoelectric currents from TM and housing[15,16].

Figure 8 shows the repeatability of two AC charge management[45] runs performed under identical conditions at 14 days interval, on July 7th and July 20th of 2015. The UV source is Experiment#2, the measurement performed with amplifier#1, $I_{LED} = 10$mA, LED duty cycle 40%. For these runs the TM potential $V(TM)$ for $V_{Bias} = 0$V was:

$$V(TM)(07/07/2015) = 247 \pm 6\text{mV}, \quad V(TM)(07/20/2015) = 227 \pm 9\text{mV} \quad (5)$$

The two charge amplifiers measure the TM potential redundantly, thus also providing consistency verification. Due to current leakage in the amplifiers, careful calibration was required to establish the zero of the TM potential.

We performed two types of tests to validate the repeatability of $V(TM)$ at $V_{Bias} = 0$V for variable photon flux intensity: a) fixed LED duty cycle and Variable Excitation Current (VEC) tests and b) fixed LED excitation current and Variable Duty Cycle (VDC) tests.

### III.b. Variable Excitation Current (VEC) Tests

To ensure the correct calibration for the dedicated PCM runs, the payload is initialized in two stages, (both with grounded housing), first with the UV-LEDs off and second at their operational current of 10 mA. The TM is then charged to +2.5 V, followed by the turn off of the LEDs and the bias. For the VEC runs, the LEDs are then turned on at currents of 1 mA, 3 mA, 4 mA, 6 mA, 7 mA, 8 mA, and 9 mA and the TM potential is measured for 45 minutes. The VEC experimental procedure is summarized in Table 1 for the test with the 9 mA LED current and 90% duty cycle. All tests in this series are similar except for the value of $I_{LED}$ used for discharging.

**Table 1. Timeline of UV discharge: $I_{LED}$ = 9 mA**

| Start Time(s) | $I_{LED}$ (mA) | $V_{BIAS}$ (V) | Operation | $V(TM)$ (mV) |
|---|---|---|---|---|
| 0 | 10 | 0 | UV ON | 232 ± 4 |
| 5 | 10 | 2.5 | Charge TM | 2,171 ± 4 |
| 14 | 0 | 2.5 | UV OFF | 1,949 ± 14 |
| 16 | 0 | 0 | Bias OFF | |
| 18 | 9 | 0 | Discharge TM | 231 ± 9 |
| 2750 | 0 | 0 | End test | |

A plot of the VEC passive discharge with LED currents from 1 mA to 9 mA (10 mA for TM charging to about 2.2 V) that follows the procedure in Table 1 is shown in Figure 9 with overlapped four runs (1 mA, 3 mA, 6 mA and 9 mA) for the first 50 s of the runs in the left panel and expanded for the 15 s to 35 s time interval for better visualization of the discharge in the right panel. For clarity, the 4 mA, 7 mA and 8 mA are not included in the figure. The data was acquired with 90% duty cycle for both the LEDs and the bias between December 8th and 27th, 2015 with Experiment#2 as source and measured with amplifier #1.

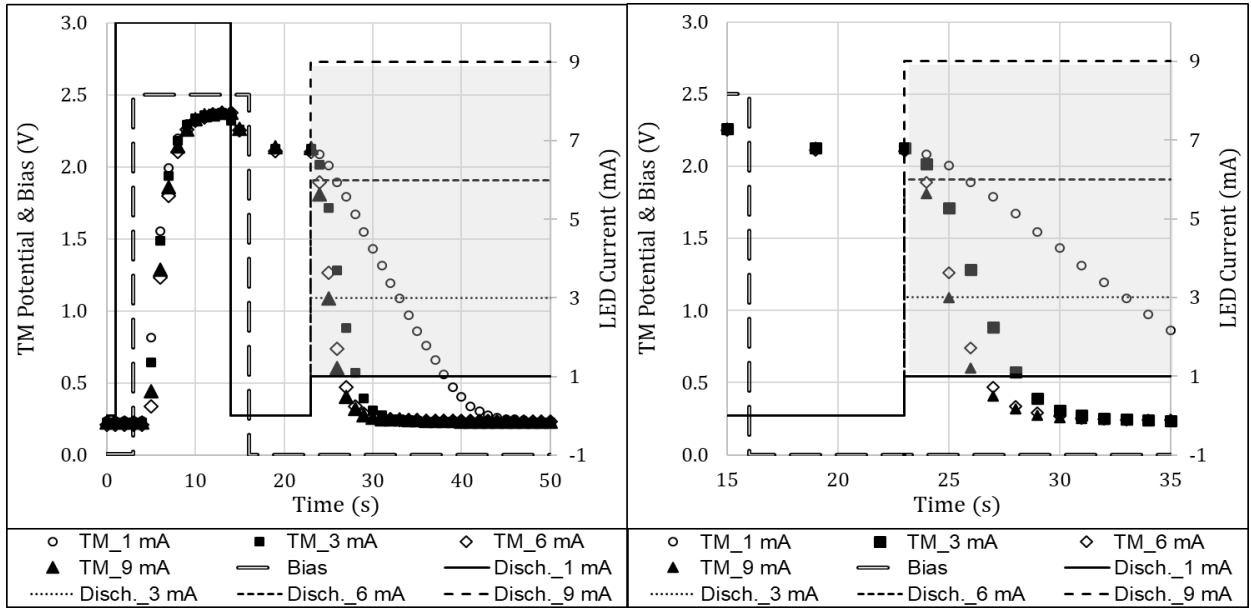

**Figure 9. PCM for VEC tests for Exp.#2, Amp.#1 (left), expanded view of discharge portion (right).**

Table 2 gives a summary of the VEC test results: LED excitation current $I_{LED}$ used to discharge the TM, test date, the TM potentials, and their errors for $V_{Bias} = 0V$ before charging to 2.5 V, $V(TM)_S$ and after the discharges $V(TM)_E$, and the differences between the readings of the two amplifiers $\Delta V(TM)_{A2}^{A1}$. Last row gives the averages and their standard deviation. The standard deviations of all $V(TM)$ values and of their averages as well as the differences between TM potentials at the start and end of each measurement, are consistent with about twice the 2.44 mV size of the digitization bin of the amplifiers' readout systems, while the standard deviation of the difference in the reading of the two amplifiers is equal to one digitization bin.

**Table 2. VEC Amp#1 $V(TM)_S$, $V(TM)_E$, $\Delta V(TM)_{A2}^{A1}$**

| $I_{LED}$ (mA) | Test Date | $V(TM)_S$ (mV) | $V(TM)_E$ (mV) | $\Delta V(TM)_{A2}^{A1}$ (mV) |
|---|---|---|---|---|
| 1 | 12/08/15 | 230 ± 4 | 226 ± 5 | 10.3 ± 2.5 |
| 3 | 12/21/15 | 227 ± 4 | 228 ± 1 | 10.4 ± 2.2 |
| 4 | 12/08/15 | 219 ± 7 | 220 ± 6 | 10.2 ± 2.6 |
| 6 | 12/26/15 | 223 ± 6 | 228 ± 6 | 10.6 ± 2.3 |
| 7 | 12/27/15 | 233 ± 2 | 230 ± 1 | 10.2 ± 2.2 |
| 8 | 12/16/15 | 235 ± 6 | 234 ± 7 | 10.5 ± 2.7 |
| 9 | 12/22/15 | 231 ± 4 | 232 ± 2 | 10.2 ± 2.4 |
| **Average** |  | **229 ± 6** | **228 ± 6** | **10.3 ± 2.4** |

Figure 10 shows the TM potential of Figure 9, as recovered after performing an inverse convolution in the time domain of the filtered data with the 4th order low-pass filter impulse response. For clarity, again only the 1 mA, 3 mA, 6 mA, 9 mA runs are shown. Figure 11 shows the data from Table 2: $V(TM)_S$ and $V(TM)_E$ in the left panel and $\Delta V(TM)_{A2}^{A1}$ in the right panel. The sigma of the scatter in the data of ≤ 6 mV corresponds to a TM residual charge repeatability of ≤ 0.17 ± 0.05 pC < $0.1 \times Q_{TM}^{max}$.

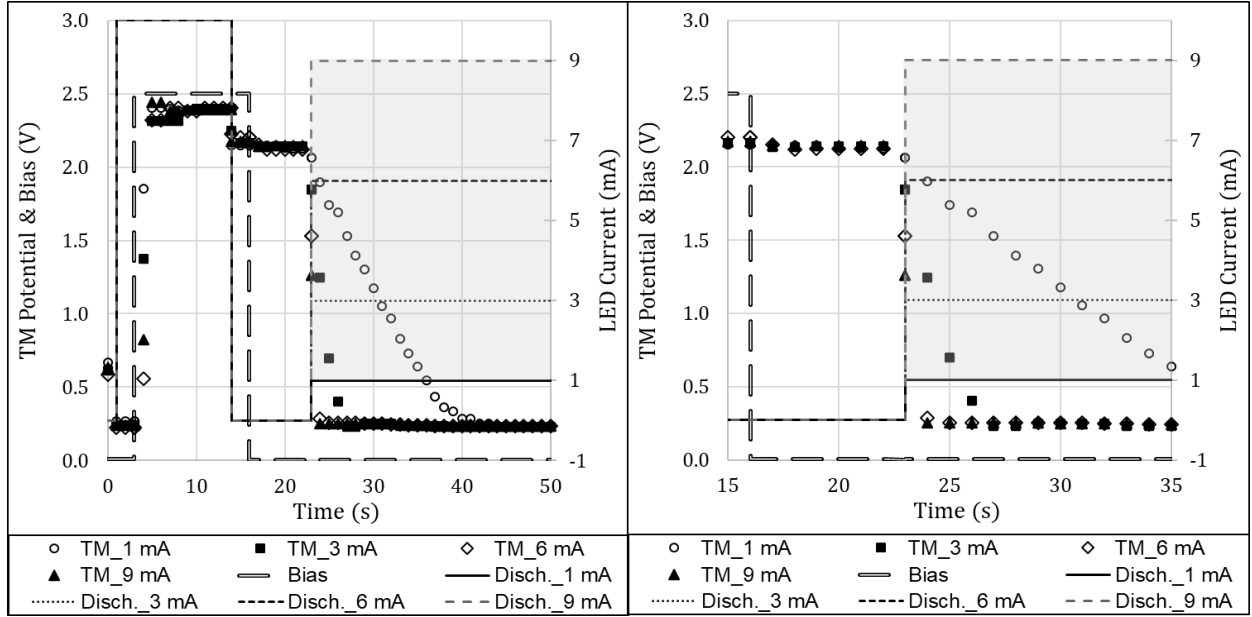

**Figure 10. Data of Figure 9 after inverse convolution for low pass 0.1 Hz filter. Exp.#2, Amp.#1**

Charge amplifier data is sampled at 1 Hz and then passed through a 4th order low-pass filter with a cut-off frequency of 0.1Hz, therefore increasing the charging and discharging time constants shown in Figure 9.

From the data of Figure 9 or Figure 10 we calculate the discharging rates by reflected photons, $dV_{TM}^R/dt$, for $I_{LED} = 1$ to 9mA, and the discharge currents, $I_{TM}^R = dQ_{TM}/dt = C(dV_{TM}^R/dt)$; see Figure 12. Note that, the sampling rate of 1 Hz causes saturation of the data as $dV_{TM}^R/dt$ approaches 1 V/s, making the measurements in this range not representative.

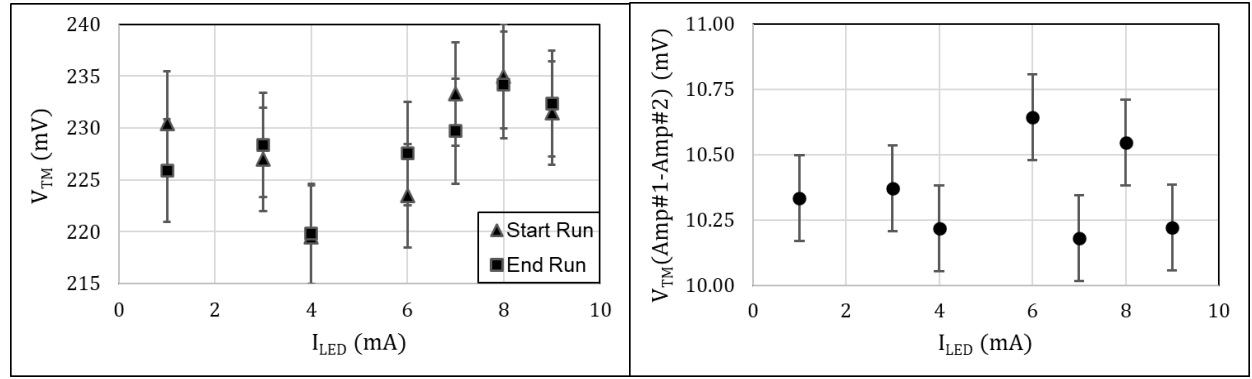

**Figure 11.** $V(TM)_S$ and $V(TM)_E$ (left), $\Delta V(TM)_{A2}^{A1}$ (right) at $V_{Bias} = 0V$ from Figure 9, 10 data

In our setup, discharging from a positive potential requires photoelectrons to accumulate on the TM, and therefore the major contribution to the photocurrent is due to the first reflection of the photon flux. $I_{TM}^R/I_{LED} = -(4.0 \pm 0.3)$ pA/mA from Figure 12, and scaling to one LED, 10 mA excitation current, and 100% duty cycle we obtain the VEC discharge current $(I_{TM}^R)^{VEC}$:

$$(I_{TM}^R)^{VEC} = I_{TM}^R/(NRD_C E_{LED}) \approx -(200 \pm 20) \text{fA/}\mu\text{W} \tag{6}$$

The ratio of the measured $I_{TM}^{VEC}$ to the nominal value $I_{TM} = (120 \pm 4)$ fA/μW from equation 2:

$$I_{TM}^{VEC}/I_{TM} \approx 1.67 \pm 0.17 \tag{7}$$

reflecting an underestimate of 50% to 85% in the product $\Psi R E_{LED}$ (photoemission efficiency, reflectivity, LED efficiency). As per equation 3, higher order reflections would only reduce the underestimate by less than 3%. We also neglect the Schottky effect [46] in the discussion of the work function, $W$, as for our applied electric fields, $E \leq 100\,\text{V/m}$, its suppression $\Delta W$ is: $\Delta W \leq 4\,\text{meV}$.

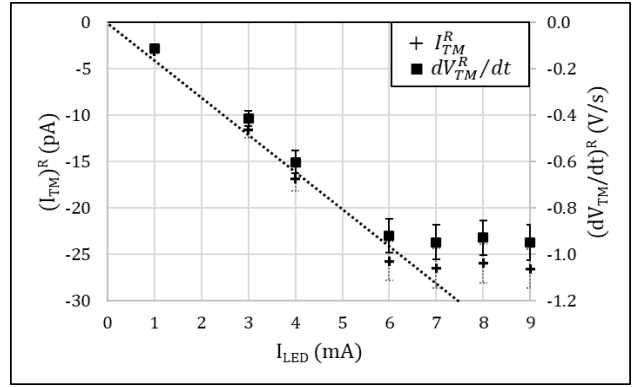

**Figure 12.** $I_{TM}^R$ and $dV_{TM}^R/dt$ for Figure 10 data.

### IIIb. Variable Duty Cycle (VDC) Tests

Variable duty cycle tests were performed with duty cycles from $D_C = 10\%$ to $D_C = 90\%$ in increments of 10% and with $I_{LED} = 10\,\text{mA}$. The bias was raised from $-2.5\,\text{V}$ to $+2.5\,\text{V}$ for 1 min at the start of each new $D_C$ setting and then lowered back to $-2.5\,\text{V}$ for 1 min, for a total run time of about 20 min. This test template was run on 11/29/2015 for Experiment #1 and on 11/22/2015 for Experiment #2. Figure 13 shows the TM potential for a VDC series source data with Experiment #1 as source, and amplifier #1 as detector on left panel, and same data deconvoluted for the 0.1 Hz low pass filter on right panel. Also shown in Figure 13 are the applied bias and, on the secondary vertical axis, the duty cycle of the LEDs.

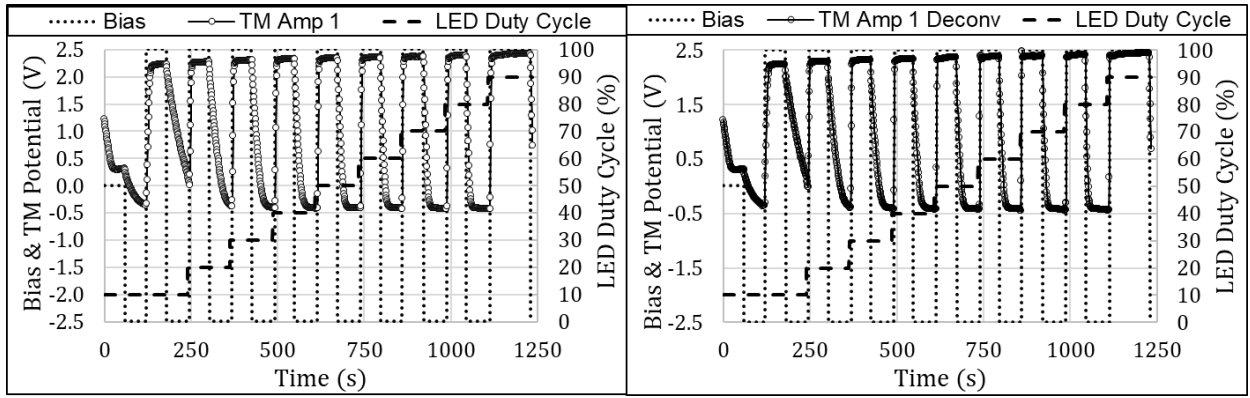

**Figure 13.** VDC test, 11/29/2015, Exp. #1, Amp. #1: left – source data, right – deconvoluted data.

The segment with $V_{Bias} = 0\,\text{V}$ is used for validation of the TM potential and shown for both VDC experiments in Table 3. We use the charging and discharging times of the VDC run on Experiment #1 to validate the photoelectron yields, and their ratio to estimate the experiment specific reflection coefficient $R$:

$$R \equiv I_{TM}^R/I_{TM}^D \qquad (8)$$

Figure 14 shows, as a function of duty cycle, the charging current by direct illumination of the TM, $I_{TM}^D$, in the left panel and the discharging current, $I_{TM}^R$, in the right panel. As for the VEC case, currents of $> 1$ pA, (or $d(V(TM))/dt > 1\,\text{V/s}$), give biased results due to the 1 Hz sampling rate. The values measured from the data in Figure 14 are:

$$I_{TM}^D = (1.08 \pm 0.08)\,\text{pA}/(\%) \quad \text{and} \quad I_{TM}^R = -(0.17 \pm 0.01)\,\text{pA}/(\%) \qquad (9)$$

We define $(I_{TM}^D)^{VDC}$ and $(I_{TM}^R)^{VDC}$ as the charging and discharging currents normalized to one LED, duty cycle $D_C = 100\%$, and $I_{LED} = 10\,\text{mA}$ excitation current:

$$(I_{TM}^D)^{VDC} = I_{TM}^R/(ND_C E_{LED}) = (85 \pm 8)\,\text{fA}/\mu\text{W} \tag{10}$$

$$(I_{TM}^R)^{VDC} = I_{TM}^R/(ND_C E_{LED}) = -(13 \pm 1)\,\text{fA}/\mu\text{W} \tag{11}$$

$$R = I_{TM}^R/I_{TM}^D = 0.16 \pm .02 \tag{12}$$

The reflection coefficient $R$ is consistent with the measurements with similar gold coatings performed pre-flight on samples[28] as well as ground tests on a different setup[15]. The ratio of measured to calculated photocurrent for the VDC tests is:

$$I_{TM}^{VDC}/I_{TM} \approx 0.71 \pm 0.07 \tag{13}$$

an overestimate of the $\Psi R E_{LED}$ product (photoemission efficiency, reflectivity, LED efficiency).

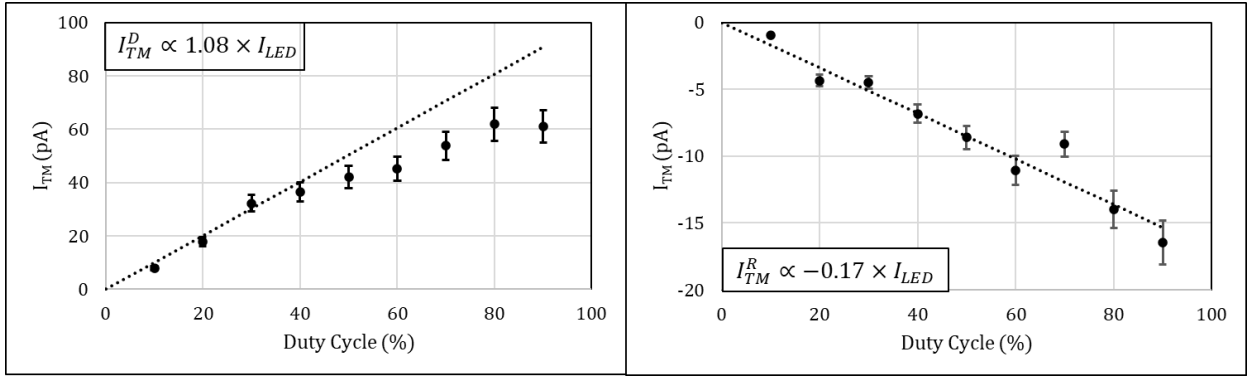

**Figure 14. Charging (left) and discharging (right) currents for deconvoluted VDC test Exp#1, Amp#1.**

However, as the reflectivity coefficient $R$ for our gold coatings is consistent over three separate experiments, both ground and space, we will attribute the non-unity results from equations 8 and 13 to the product $\Psi E_{LED}$ (photoemission and LED efficiencies) differences between Experiment#1 and Experiment#2. Using the calibration PD data in Figure 7 we assume that the main difference is in all likelihood due to the actual LED efficiency, $E_{LED}$, as it can vary significantly between LED's, with their particular mounting arrangements, and with the variation in the optical paths of the UV photons between the two experiments. Furthermore, the photoemission efficiency $\Psi$, is constant between measurements but its exact value is not determined for these tests.

TM potential values at zero-bias, $V_{Bias} = 0$V, acquired during all the various space-based charge measurements, under a variety of conditions (including the VEC and VDC tests), and over a period of six months are shown in Table 3, and result in an average TM potential for the two experiments and the two amplifiers of:

$$\langle V_{A1}^{E1} \rangle = (312 \pm 6)\,\text{mV} \quad \langle V_{A2}^{E1} \rangle = (326 \pm 5)\,\text{mV} \tag{14}$$

$$\langle V_{A1}^{E2} \rangle = (228 \pm 6)\,\text{mV} \quad \langle V_{A2}^{E2} \rangle = (216 \pm 6)\,\text{mV} \tag{15}$$

where the subscript denotes the amplifier number and the superscript the experiment number.

Averages and their errors in equations 14 and 15 are computed using their average values $\langle V_{A1} \rangle$ and $\langle V_{A2} \rangle$ and the standard deviation for each value $\sigma \langle V_{A1} \rangle$ and $\sigma \langle V_{A2} \rangle$ from Table 3. The errors in equations 14 and 15 are consistent with the averages of the standard deviations for each measurement in Table 3 and equal to $\cong (6 \pm 1)$mV, and equal to about twice the 2.44 mV digitization minimum. See also Table 2 and Figure 8, Figure 9, Figure 10 Figure 11for consistency with the start and end values of the TM potential at zero bias for the VEC tests.

**Table 3.** Summary of all TM potentials for the space-based tests with $V_{Bias} = 0V$

| Date | Test # | Exp # | Duty cycle (%) | $I_{LED}$ (mA) | $\langle V_{A1} \rangle$ (mV) | $\sigma\langle V_{A1} \rangle$ (mV) | $\langle V_{A2} \rangle$ (mV) | $\sigma\langle V_{A2} \rangle$ (mV) | $\langle V_{A1} - V_{A2} \rangle$ (mV) |
|---|---|---|---|---|---|---|---|---|---|
| 12/9/2014 | 1 | 2 | 60 | 10 | 270 | 14 | 263 | 14 | 7 |
| 7/1/2015 | 2 | 2 | 40 | 10 | 247 | 6 | 234 | 5 | 12 |
| 7/13/2015 | 3 | 2 | 50 | 10 | 227 | 8 | 214 | 8 | 12 |
| 7/14/2015 | 4 | 2 | 50 | 10 | 229 | 9 | 217 | 9 | 13 |
| 7/19/2015 | 5 | 2 | 40 | 10 | 227 | 9 | 216 | 9 | 10 |
| 7/21/2015 | 6 | 2 | 75 | 10 | 222 | 8 | 212 | 8 | 11 |
| 8/3/2015 | 7 | 1 | 50 | 10 | 307 | 5 | 320 | 6 | -13 |
| 8/5/2015 | 8 | 1 | 50 | 10 | 314 | 5 | 326 | 6 | -12 |
| 8/13/2015 | 9 | 1 | 40 | 10 | 318 | 10 | 330 | 12 | -12 |
| 8/16/2015 | 10 | 1 | 75 | 10 | 317 | 4 | 329 | 4 | -12 |
| 8/25/2015 | 11 | 1 | 90 | 10 | 314 | 4 | 328 | 3 | -15 |
| 8/29/2015 | 12 | 2 | 90 | 10 | 231 | 3 | 219 | 4 | 12 |
| 9/2/2015 | 13 | 1 | 90 | 10 | 315 | 3 | 329 | 3 | -14 |
| 9/3/2015 | 14 | 2 | 90 | 10 | 223 | 8 | 212 | 7 | 11 |
| 11/22/15 | 15 | 2 | 10 | 10 | 240 | 3 | 225 | 3 | 13 |
| 11/27/2015 | 16 | 2 | 90 | 10 | 233 | 5 | 221 | 5 | 12 |
| 11/29/2015 | 17 | 1 | 10 | 10 | 303 | 7 | 317 | 8 | -15 |
| 12/8/2015 | 18 | 2 | 90 | 10 | 231 | 5 | 217 | 5 | 14 |
| 12/8/2015 | 19 | 2 | 90 | 1 | 216 | 12 | 206 | 11 | 10 |
| 12/8/2015 | 20 | 2 | 90 | 10 | 223 | 7 | 208 | 7 | 15 |
| 12/8/2015 | 21 | 2 | 90 | 4 | 220 | 11 | 210 | 11 | 10 |
| 12/16/2015 | 22 | 2 | 90 | 10 | 244 | 6 | 231 | 6 | 14 |
| 12/16/2015 | 23 | 2 | 90 | 8 | 232 | 9 | 221 | 9 | 10 |
| 12/21/2015 | 24 | 2 | 90 | 10 | 227 | 4 | 214 | 4 | 13 |
| 12/21/2015 | 25 | 2 | 90 | 3 | 220 | 9 | 210 | 9 | 10 |
| 12/22/2015 | 26 | 2 | 90 | 10 | 231 | 3 | 217 | 5 | 14 |
| 12/22/2015 | 27 | 2 | 90 | 9 | 231 | 9 | 221 | 10 | 10 |
| 12/26/2015 | 28 | 2 | 90 | 10 | 222 | 7 | 208 | 6 | 14 |
| 12/26/2015 | 29 | 2 | 90 | 6 | 230 | 6 | 219 | 6 | 11 |
| 12/27/2015 | 30 | 2 | 90 | 10 | 233 | 3 | 220 | 3 | 13 |
| 12/27/2015 | 31 | 2 | 90 | 7 | 232 | 6 | 222 | 6 | 10 |

Due to the difference in illumination of the TM by Experiments #1 and #2 the average of all readings of Amplifier #1 and Amplifier #2 at $V_{Bias} = 0V$ differ for the two experiments by (see Table 3):

$$\langle V_{A1}^{E1} - V_{A2}^{E1} \rangle = -13.1 \pm 1.2 \text{ mV} \quad \langle V_{A1}^{E2} - V_{A2}^{E2} \rangle = 11.7 \pm 1.8 \text{ mV} \tag{16}$$

The opposite signs of the difference in the readings for $V(TM)$ between amplifiers #1 and #2 for all Experiments #1 and #2 tests, as well as their statistical consistency, indicate that the charge distribution on the TM is dependent on the geometry of the illumination by the photoemission source and their average error consistent with the 2.44 mV digitization bin. Note, that calibration differences between amplifiers would result in the same sign for the difference between their readings.

Figure 15 shows all flight data with zero bias, the Table 3 data, as function of date on left panel and, for clarity, equally spaced in chronological order (as function of test number) on right panel.

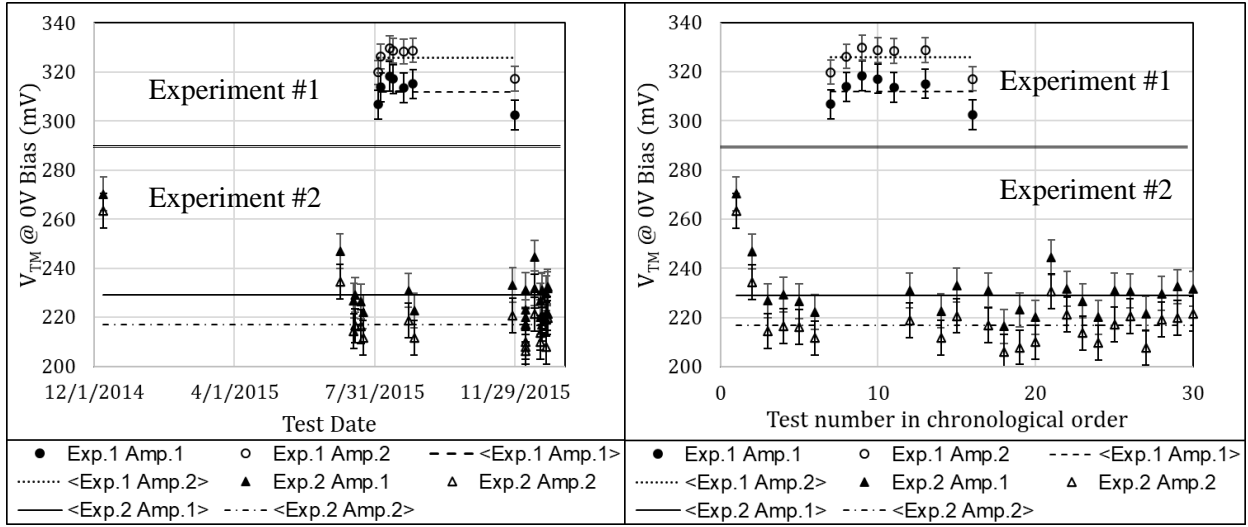

**Figure 15. Flight zero bias data arranged by date (left) and equally spaced in chronological order (right)**

Averages for each amplifier of each of Experiments #1 and #2 are also shown. Outgassing times in excess of 6 months and the flooding of the system with UV contribute to the stabilization of the equilibrium TM potentials at zero bias; see first two measurements with Experiment#1.

## IV. Ground tests

Two series of ground tests with relevance to this work were performed as part of a larger study[15] using a re-configured flight backup unit for the UV-LED SaudiSat 4 mission[26].

For the first series, S1 on 9/10-9/11/2020, the changes to the unit were the use of only one 255 nm LED of TO39 flat window type placed in the center of the $x$ bias plate, the suspension of the spherical TM by a 1 mm gold wire, and the removal of the $z$, and $-z$ TM supporting structures: see Figure 16. Four levels of LED excitation current, $I_{LED}$, were used, 1 mA, 2mA, 5 mA, and two separate runs at 10 mA. The duty cycle was 100%.

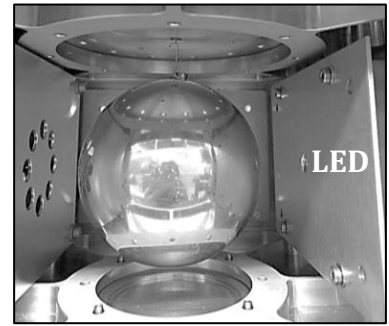

**Figure 16. S1 ground test system**

**Table 4. Potentials for S1 series**

| Test Date | $I_{LED}$ (mA) | $V^{S1}$ (mV) | $\sigma V^{S1}$ (mV) |
|---|---|---|---|
| 9/11/20 | 1 | 627 | 10 |
| 9/11/20 | 2 | 620 | 9 |
| 9/11/20 | 5 | 628 | 5 |
| 9/11/20 | 10 | 629 | 3 |
| 9/10/20 | 10 | 632 | 4 |
| **Average** | | **627±4** | **6±2** |

Table 4 gives the TM potentials, $V^{S1}$, and their standard deviations, $\sigma V^{S1}$, for the four runs, with their averages in the last row. $V^{S1}$ is independent of the LED excitation current $I_{LED}$, though it very clearly changed with the modifications in configuration from the space-based tests.

The unweighted standard deviation of the $V^{S1}$ measurements is consistent with the average of $\sigma V^{S1}$, where the $\sigma V^{S1}$ values are the result of the scattering of points in each $V^{S1}$ measurement.

Figure 17 shows $V^{S1}$, the TM potentials data for S1 series ground-test from Table 4, with their corresponding errors $\sigma V^{S1}$, as well as the average $\sigma V^{S1}$ value and its error band.

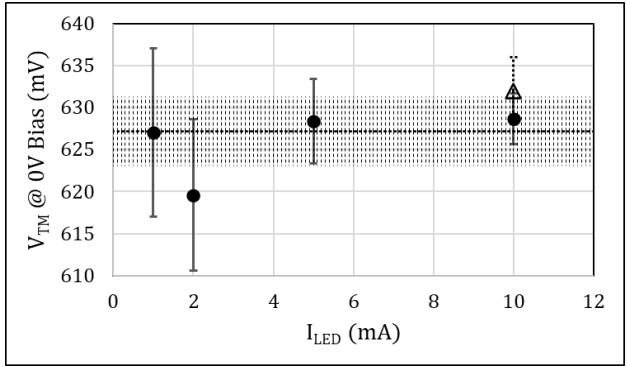

**Figure 17. S1 data from Table 4.**

For the second series, S2 on 9/4-10/8/2020, we used the same LED as in S1; 255 nm wavelength, TO39 flat window type, and placed in the center of the $x$ bias plate. The spherical TM was suspended as in the S1 configuration for test#1, rotated 180 degrees in test#2, and resting on an Ultem™ resin for test#3. Figure 18 shows the configuration for test#3 that included a secondary UV diode, LED2, that can illuminate the bias plate, and that was used for a different experiment. For the S2 series all housing plates, except the $x$ plate, were removed, the duty cycle was 100%, and the excitation current $I_{LED} = 10mA$.

S2 tests #1 and #2 check for differences in the photoemission coefficients $R$ and $\Psi$ (reflectivity and photoemission efficiency) between the gold coatings on opposing hemispheres of the TM. Test #3 checks for variation in the TM potential equilibrium for small changes in the configuration of the TM enclosure.

Table 5 gives the test numbers, date, TM potentials, $V^{S2}$, and their standard deviations, $\sigma V^{S2}$, calculated from the scattering of their data points. We note that the gold coatings on the opposing faces of the TM have equal photoemission coefficients within one standard deviation even after the venting and re-pumping of the system and the reorientation of the sphere. The change in TM equilibrium potential between the suspended and the Ultem supported TM cannot confidently be completely attributed to only the modification in system geometry, as the entire system was vented, cleaned, and re-configured after test #3; even though somewhat similar operations did not change the results between tests #1 and #2. Figure 19 shows the data, $V^{S2}$ and $\sigma V^{S2}$ from Table 5.

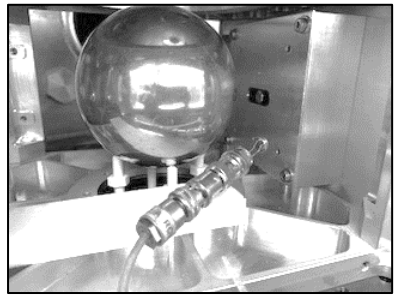

**Figure 18. S2 ground test system**

**Table 5. TM potentials for S2 series**

| Test # | Date | TM Configuration | $V^{S2}$ (mV) | $\sigma V^{S2}$ (mV) |
|---|---|---|---|---|
| 1 | 10/1/20 | Suspended | 547 | 3 |
| 2 | 10/8/20 | Suspended 180° | 554 | 6 |
| 3 | 9/4/20 | Ultem support | 528 | 6 |

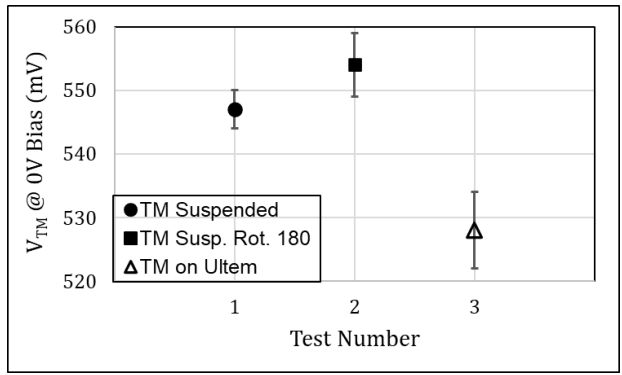

**Figure 19. S2 data from Table 5**

## Conclusions

Space and ground based acquired data demonstrate that, for photoemission for a UV-LED source in a stable system configuration, the equilibrium TM potential at zero bias is constant to less than ± 6 mV (170 fC in our 28 pF system). 'Stable system configuration' requires venting of the unit to space, in a 600 km orbit, for up to 6 months as well as conditioning the TM and TM housing surfaces by exposing them to UV. The space experiments showed this level of reproducibility for the entire duration of the 6 months 'stable system configuration' testing, for UV photon fluxes varying by a factor of 10 and TM potentials approaching equilibrium from both positive and negative voltages of up to 2.5 V. Ground tests validated this level of stability for the TM equilibrium potential, as well as its independence of UV photon flux variations by a factor of 10, albeit over a much shorter durations for the set of measurements.

Reproducibility of the TM potential at zero bias indicates the stability of the LED and the photoemission efficiencies and of the UV reflectivity for gold coatings deposited by the same process. The equal TM potentials, obtained after rotating the sphere by 180 degrees (ground test S2, tests #1 and #2), indicate that the photoemission properties of the gold coatings are reproducible over their entire area.

The demonstrated reproducibility of the photoemission properties is a necessary condition for the implementation of robust passive (or self-adjusting) charge control of floating TMs using either the slow or fast photoelectron approaches[15,16]. These techniques have the following advantages: a) Infrequent or no charge measurement required, b) No stable power of the UV illumination necessary, c) No precise duration of the UV illumination necessary, d) No need for the accurate a priori determination of the photoemission properties of the surfaces, e) No complex, precise and or repeatable processing and maintenance of the UV illuminated surfaces required and f) Simple in-flight fine-tuning for unforeseen events of the passive fast photoelectrons charge management system. The passive slow photoelectrons charge management system is self-tuning[15].

## Acknowledgements

This work was made possible by grants and support provided by the National Air and Space Administration through the Ames Research Center, the King Abdulaziz City for Science and Technology in Saudi Arabia, and SN&N Electronics, Inc. of California. We are indebted to the W.W. Hansen Experimental Physics Laboratory at Stanford University for making available facilities and equipment critical to this work.